\def\expec#1{\langle#1\rangle}

\def\beq#1{\begin{equation}\label{#1}}
\def\eeq{\end{equation}}
\def\beqa#1{\begin{eqnarray}\label{#1}}
\def\eeqa{\end{eqnarray}}

\def\spose#1{\hbox to 0pt{#1\hss}}
\def\simlt{\mathrel{\spose{\lower 3pt\hbox{$\mathchar"218$}}
     \raise 2.0pt\hbox{$\mathchar"13C$}}}
\def\simgt{\mathrel{\spose{\lower 3pt\hbox{$\mathchar"218$}}
     \raise 2.0pt\hbox{$\mathchar"13E$}}}
\def\simpropto{\mathrel{\spose{\lower 3pt\hbox{$\mathchar"218$}}
     \raise 2.0pt\hbox{$\propto$}}}

\def\ed{\end{document}}

\def\u{{\bf u}}

\def\o{{\bf o}}

\def\th{\hbox{\boldmath $\theta$}}

\def\gmp{\gamma_{+}}
\def\gmc{\gamma_{\times}}

\newcommand{\lexp}{\mathop{\langle}}    
\newcommand{\rexp}{\mathop{\rangle}}    

\def\fun#1#2{\lower3.6pt\vbox{\baselineskip0pt\lineskip.9pt
        \ialign{$\mathsurround=0pt#1\hfill##\hfil$\crcr#2\crcr\sim\crcr}}}

\def\tlg{\tilde\gamma}

\def\exp#1{\lexp #1 \rexp} 
\documentstyle[emulateapj,danonecolfloat]{article}
\def\NoApjSectionMarkInTitle#1{#1.\ }
\begin{document}
\twocolumn[


\journalid{337}{15 January 1989}
\articleid{11}{14}

\submitted{\today. To be submitted to ApJ.}

\title{Higher-order moments of the cosmic shear and other spin-2 fields}

\author{Matias Zaldarriaga$^{1,2}$ \& Rom\'an Scoccimarro$^{1}$}

\affil{$^{1}$Center for Cosmology and Particle Physics, Physics Department, \\ 
New York University, 4 Washington Place, New York, NY 10003}

\affil{$^{2}$Institute for Advanced Study, Einstein  Drive, Princeton, NJ 08540}

\vskip 1pc

\begin{abstract}
We present a method for defining higher-order moments of a spin-2
field on the sky using the transformation properties of these
statistics under rotation and parity. For the three-point function of the cosmic shear
we show that the eight logically possible combinations of the shear in three points
can be divided into two classes, four combinations are even under parity
transformations and four are odd. We compute the
expected value of the even parity ones in the non-linear regime using the halo model 
 and conclude that on small scales of the four combinations there is one that
is expected to carry most of the signal for triangles close to
isosceles. On the other hand, for collapsed triangles all four
combinations are expected to have roughly the same level of signal
although some of the combinations are negative and others positive. We
estimate that a survey of a few square degrees area is enough to detect this
signal above the noise at arc minute scales.
\end{abstract}

\keywords{cosmology: theory -- gravitational lensing -- large scale structure
of universe-- cosmic microwave background }

]


\section{Introduction}
\label{introduction}

Gravitational lensing is quickly becoming a major tool for doing
cosmology. Mass determinations of cluster of galaxies are now routine,
there are several detections of weak lensing by the large scale
structure of the universe and measurements of galaxy-galaxy
lensing. Detections of weak lensing (Bacon et al. 2000; Kaiser et
al. 2000; Van Waerbeke et al. 2000; Wittman et al. 2000) are
particularly important as they will provide constraints on the matter
budget of the universe and an independent determination of the power
spectrum of the dark matter fluctuations over a wide range of scales
that can be then compared with other measurements such as those 
from galaxy clustering or the anisotropies in the Cosmic Microwave
Background (CMB).  

As a result of non-linear gravitational evolution, the projected mass
density field on the sky ($\kappa$), which is responsible for the
deflections that lead to the measured shear, is expected to be highly
non-Gaussian even if the initial seeds of density perturbations were
perfectly Gaussian.  The aim of observations that attempt to make
projected mass maps is not only to measure the projected mass power
spectrum, but also higher-order moments which contain additional
information.  It has been stressed that valuable constraints on
cosmological parameters are expected to come from a joint measurement
of the variance and higher-order moments such as the skewness of the
weak lensing convergence field because the variance is strongly
dependent on both the amplitude of the mass power spectrum and the
density of the Universe, while the skewness (properly normalized) is
essentially a measure of the latter (Bernardeau, Van Waerbeke \&
Mellier 1997; Jain \& Seljak 1997; Schneider et al. 1998; Van
Waerbeke, Bernardeau \& Mellier 1999; Van Waerbeke et al. 2000).

The non-linear growth of structure is very important on the length
scales relevant for weak lensing, so estimates for the level of
non-Gaussianity in $\kappa$ come from two different techniques:
N-body simulations (Couchman, Barber \& Thomas 1999; Jain, Seljak \&
White 2000; White \& Hu 2000) and semi-analytic models (Jain \& Seljak
1997; Schneider et al. 1998; Van Waerbeke et al 2001).  N-body
simulations have been used to create mock $\kappa$ maps from which
higher-order moments have been measured. Semi-analytic techniques such
as the halo model (see Cooray \& Sheth 2002 for a recent review) or
proposals about the behavior of higher-order correlations in the
non-linear regime (Scoccimarro \& Frieman 1999) have also been used to
make weak lensing predictions (Hui 1999; Cooray, Hu \&
Miralda-Escud\'e 2000; Cooray \& Hu 2001).

So far most of these theoretical approaches have dealt with $\kappa$;
however the quantity that is directly measured is the shear
$\gamma$\footnote{What is actually measured is the reduced shear
$\gamma/(1-\kappa)$, thus on sufficiently small scales this may change
theoretical predictions. The spin properties of the three-point
function we will discuss apply both to the shear and the reduced
shear}. The crucial difference between $\kappa$ and $\gamma$ is that
the shear is a spin-2 field on the sky while $\kappa$ is just a
scalar. Thus the shear has two components at each position on the sky
so when designing any statistics for the shear one has to make sure
that the statistic does not depend on some arbitrary choice of
coordinate system.  Moreover the transformation from shear to $\kappa$
cannot be done exactly if one only has a shear map over a finite
region on the sky and only at the position of the background galaxies.
To circumvent this issue, measures of the shear at different points on
the sky can be combined to form statistics that are invariant under
rotation and can be expressed as weighted averages of $\kappa$; the
best example being the aperture mass (\cite{kaiser94,sch98}). We will
show that the amplitude of the three-point function varies with
triangle configuration and can be both positive or negative. As a
result not all combinations of the three-point function are optimal
from a signal to noise perspective. 

A natural way to look for non-Gaussianity is then to look at
statistics of the aperture mass. In this paper we will focus on
another approach, directly defining higher-order statistics in terms
of the shear field that are independent of the coordinate system.  Any
statistic of the shear can ultimately be written as a linear
combination of the statistics we present in this paper. We will show
however that not all the possible higher-order moments are expected to
have the same level of cosmological signal and, moreover, the sign of
these statistics can be positive or negative depending on the
configuration of the points. The advantage of our approach is that it
will allow us to isolate the configurations that have most
cosmological signal and avoid suppression of signal to noise in the
measurements or cancellations that would result from an arbitrary
linear combination.

A first attempt has been made to define a three-point function for
the shear field in Bernardeau et al (2002a). Their prescription
corresponds to integrating over a particular linear combination of the
four different shear three-point functions that we define in this
work. They have applied their statistic to a cosmic shear survey and
reported a detection at approximately $5\sigma$ level (Bernardeau et
al. 2002b). Given these encouraging results it is worth to consider in
more detail how to construct shear three-point functions and what is
expected theoretically about their order of magnitude and sign
depending on the particular configuration of the points.

This paper will be written using weak lensing language, but identical
issues arise if one wants to define higher-order moments of the CMB
polarization field. The analogue of the shear components are the $Q$
and $U$ Stokes parameters and the analogue of the projected mass
density is usually called the $E$ field. Even if the initial
conditions are Gaussian, higher-order moments of the CMB polarization
can be generated by secondary effects such as lensing so the
statistics presented here will be equally applicable for the CMB.

\section{Defining higher-order moments for spin-2 fields}

In this section we will show how to define higher-order moments of the
weak lensing shear or CMB polarization fields in a way that is
geometrically meaningful. The problem is that the shear is a spin-2
field and thus at each point on the sky it has two components.  Just
as in the case of a vector field the values of these components depend
on the choice of coordinate system. 
If at any given point one rotates the coordinate system
used to define the shear components by an angle $\alpha$
(anticlockwise in our convention) the shear field has to be
transformed as, 
\beqa{transfshear} \gamma_1^\prime&=& \ \ \cos
2\alpha\ \gamma_1 + \sin 2\alpha \ \gamma_2 \nonumber \\
\gamma_2^\prime&=& - \sin 2\alpha\ \gamma_1 + \cos 2\alpha \ \gamma_2
\eeqa

Any meaningful statistic of the shear measured in a set of points has
to depend only on the distances and relative orientation of these
points and not on their absolute position on the sky or their
orientation with respect to some fiducial origin. For example the two
point function has to depend only on the distance between the two
points and the three-point function only on the size and shape of the
triangle formed by the three points.

The way to overcome this problem of definition for the second moment
is well known, one uses the separation vector to define a ``natural"
coordinate system. The idea is to align the coordinate system so that
one of the axis lies on the great circle joining the two points and
uses that coordinate system to define the two components of the shear
(\cite{miral91,kaiser92,kamion}).

In general the way to define higher-order moments that are invariant
under rotation is by contracting the measured shear at the three
points with suitable combinations of the vectors that define the sides
of the triangle, are invariant under translation and have the correct spin to 
compensate for the spin of the shear. Clearly this procedure is not unique 
as there are many such combinations.  In this paper we will propose 
a simple, intuitive but at the same time
geometrically meaningful way to define higher-order moments. 

\begin{figure}[tb]
\centerline{\epsfxsize=10cm\epsffile{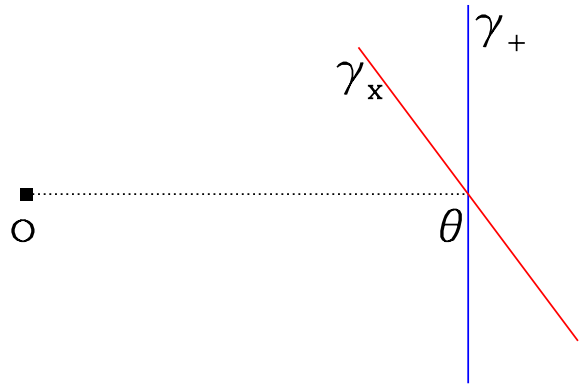}}
\caption{The point $\o$ is used to define the $\gmp$ and $\gmc$
components of the shear at point $\th$. The two rods indicate the
ellipticity at $\th$ that would produce a positive $\gmp$ and $\gmc$ }
\label{defgm}
\end{figure}

An N-point function is characterized by the set of points $X=\{\th_i
\}$ ($i=1,\ \dots \ N$) where the shear or the CMB polarization is
measured. We denote $2D$ vectors on the sky with boldface. We then
define the ``center of mass" of $X$, 
\beq{com} \o={1\over N}
\sum_{i=1}^N \th_i 
\eeq 
and use $\o$ as the origin when defining the shear at each of the
points in $X$. At every point we can define a component of the shear
along the direction that separates $\o$ and $\th_i$ which we can call
$\gmp$ and a component which is measured in the coordinate system that
is rotated by $45^o$, which we call $\gmc$. Figure \ref{defgm}
illustrates how the $\gmp$ and $\gmc$ are defined.  This is totally
analogous to the procedure for the two point function, but now the two
points used in the definition of $\gmp$ and $\gmc$ are $\o$ and
$\th_i$ instead of the two points in the two point function. The key
to our procedure is that by defining an origin based only on the
points in $X$ we make sure that our statistic depends only on
intrinsic properties of $X$ and not on a fiducial origin.

Another important property of spin-2 fields is that they can be
decomposed into two scalar potentials, one that is even under parity
($E$) and one that is odd ($B$)
(\cite{kamion,selzal,crit01,sch02}). In fact for finite sky coverage
or for maps with holes, there is a third family of modes, the
ambiguous modes for which one does not have enough information to
decide whether the contribution is coming from $E$ or from $B$
(\cite{lewis01,bunn02}). Weak gravitational lensing only produces $E$
modes ($E$ is nothing but the projected mass density $\kappa$).

The properties under parity transformation of $\gmp$ and $\gmc$ are
different. As is clear from the definition, to obtain $\gmc$ one has
to rotate the coordinate system anticlockwise. This rotation changes
direction when we do a parity transformation. This means that under
parity $\gmp^\prime \rightarrow \gmp$ but $\gmc^\prime \rightarrow
-\gmc$.

\begin{figure}[tb]
\centerline{\epsfxsize=9cm\epsffile{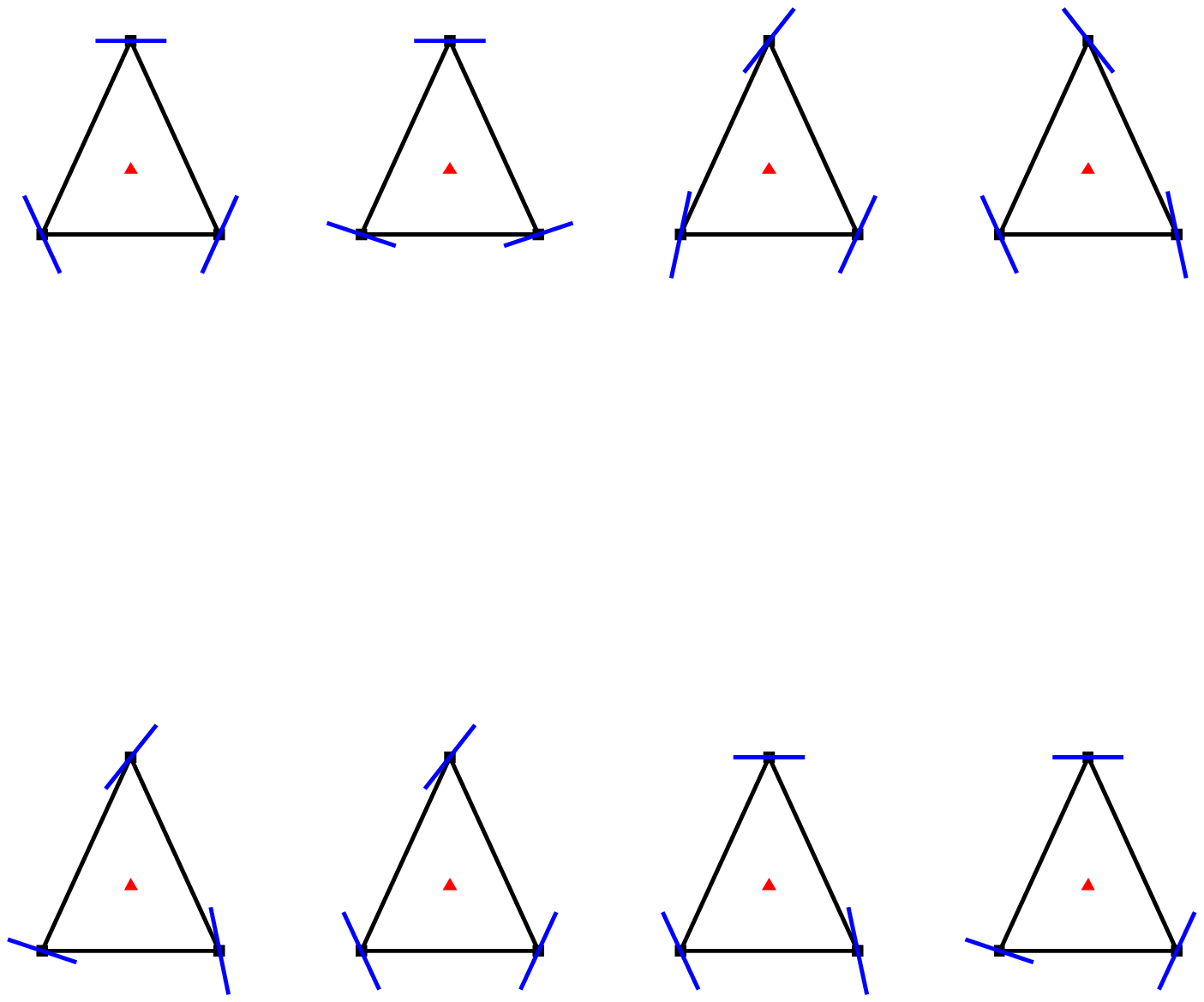}}
\caption{The top row (from left to right) shows four shear patterns that would produce a
positive value for $\exp{\gmp\gmp\gmp}$, $\exp{\gmc\gmc\gmp}$,
$\exp{\gmc\gmp\gmc}$, $\exp{\gmp\gmc\gmc}$.  The bottom row shows four
shear patterns that would produce a positive value for
$\exp{\gmc\gmc\gmc}$, $\exp{\gmp\gmp\gmc}$, $\exp{\gmp\gmc\gmp}$,
$\exp{\gmc\gmp\gmp}$} .
\label{poscorr}
\end{figure}

As a consequence of the difference in behavior of $\gmp$ and $\gmc$
any estimator that contains an odd number of $\gmc$ is odd under
parity. The difference in their parity behavior will make some of 
the three-point functions vanish. To illustrate what this means we can
consider the case of the three-point function for equilateral
triangles. In principle there are 8 different combinations
of the two shear components at the three points, however only four of
them: $\exp{\gmp\gmp\gmp}$, $\exp{\gmp\gmc\gmc}$,
$\exp{\gmc\gmp\gmc}$, $\exp{\gmc\gmc\gmp}$ contain any signal from
weak lensing because they are the only four that are even under
parity.  We show patterns that produce positive values for these
correlations in Fig.~\ref{poscorr}. For a discussion on how
the difference in behavior under parity of the three-point functions of
general configurations constraints their possible values see \cite{takabhuv}.

Being able to separate combinations that have signal from those that
do not is very important not to dilute the signal one is trying to
measure when combining or binning  the measured statistics to obtain a
detection.  Moreover for equilateral triangles for example 
the four odd combinations: $\exp{\gmc\gmc\gmc}$,
$\exp{\gmp\gmp\gmc}$, $\exp{\gmc\gmp\gmp}$, $\exp{\gmp\gmc\gmp}$ may
prove useful to monitor systematic problems in the data or to identify
interesting physical effects such as clustering of the background
sources, the effects of multiple scatterings, etc. 
(\cite{bernardeau98,jsw00}).

Any connected N-point function of the shear can be written as a linear
combination of the connected N-point functions of the two potential
$E$ and $B$. It is worth noting that in the case of connected N-point
functions with odd number of legs there will be some configurations 
(such as those where the lenghts of all sides are equal) that can be 
separated into a set that only depends on the higher-order
correlations of $E$ and a set
that only depends on the higher moments of $B$ (assuming $E$ and $B$
are independent). We can argue this only on the basis of their
bahavir under parity. The product of an odd number of $E$'s is even
under parity while a product of an odd number of $B$'s is odd, thus 
the even parity N-point functions of the shear receive
contributions only from $E$ and the odd ones only from $B$.  For an N-point
function with an even number of
legs this is not true because the product of an even number of 
$E$s or $B$s is even under parity. The clearest example of this is the
two-point correlation
function where $\exp{\gmp\gmp}$ and $\exp{\gmc\gmc}$ receive
contributions from both $\expec{EE}$ and $\expec{BB}$ while
$\exp{\gmc\gmp}$ is zero if there is no $E-B$ cross-correlation.
 
As we mentioned above the way to make a three-point function (or an N-point function 
for that matter) that is scalar is to contract the shear at the three points with some 
combination of the  vectors that form the sides that transforms appropriately under rotations
to cancel the spin of shear. That is we need to construct spin$-2$ combinations of these 
vectors.   Our proposed scheme is easy to understand. For each member of $X$
we define $\tilde{\th}=\th -\o$ and we construct two spin$-2$ quantities,
\beqa{Pdef}
{\bf \bar P}_{+} &=&(\tilde \theta^2_x-\tilde \theta^2_y, 2 \tilde \theta_x \tilde \theta_y)/
 \tilde \theta^2 \nonumber \\
&=& (\cos 2\tilde \phi, \sin 2 \tilde \phi)\nonumber \\
{\bf \bar P}_{\times} &=&(-2 \tilde \theta_x \tilde \theta_y,
\tilde \theta^2_x-\tilde \theta^2_y)/ \tilde \theta^2 \nonumber \\
&=& (-\sin2 \tilde \phi, \cos2\tilde \phi).
\eeqa
The statistics we proposed are obtained by contracting the above quantities with the 
shear three-point function. For example,
\beq{example}
\expec{\gmp\gmp\gmp}= {\bar P}_{+}^{\mu_1}{\bar P}_{+}^{\mu_2} \bar{P}_{+}^{\mu_3}\expec{\gamma_{\mu_1}\gamma_{\mu_2}\gamma_{\mu_3}},
\eeq
where the index $\mu$ runs over the two components of the shear. The other 
three-point functions that we define are obtained by replacing some of the $\bar{P}_{+}$ by 
$\bar{P}_{\times}$. Finally we note that  the vector $\tilde{\th}_1$ 
is nothing but $\tilde{\th}_1= [(\th_1-\th_2)+(\th_1-\th_3)]/3$, i.e. basically 
the sum of the vectors that define the sides of the triangle that cross at $\th_1$.  The
same is true for the other vertices.

\section{Worked Examples}

Our objective in this section is to get some intuition into how these
higher-order correlations behave. To keep things simple we will just
focus on the three-point function. To get some idea of how these
functions behave we will work in the context of the halo model,
i.e. the dark matter is assumed to be distributed in a collection of
halos of different mass. Although analytic approximations and fits to
numerical simulations exist for the profile of these halos and their
mass function, this modeling is clearly simplistic to model the shear
field as in reality halos are not spherical. This particularly is
bound to affect the configuration dependence of the three-point
function. Thus definite predictions will need more detailed modeling
and comparison with direct measurements using numerical simulations.

Our aim in this paper is more modest, we only want to gain some
insight into how these different three-point functions behave, if they
are positive or negative for example. We will start by considering the
simple case of lensing by a singular isothermal sphere. We do this
because the calculation of the three-point function can be done
analytically, and it provides a useful check on the numerical code used 
to do the same calculations in the halo model, presented in 3.2. 

\subsection{ The Singular Isothermal Sphere}

In this section we will calculate the three-point function that
results from a single halo with a power-law density run. For definiteness we
will write down formulas for a singular isothermal sphere (SIS), but
other power laws can be calculated in analogous way.  For a SIS the
density $\rho(r)$ depends on distance from the center $r$ as
$\rho(r)\propto r^{-2}$; therefore, both the projected mass density
$\kappa$ and the shear $\gamma$ scale with projected separation as
$r_\perp^{-1}$.

The shear pattern around a spherical halo is tangential centered at
the origin of the halo which we will call $\u$.  When defining the
shear components at a point $\th$, we will need to rotate the shear
elements so as to define them relative to the vector $\th-\o$. We will
call this angle of rotation $\alpha$. We can write
\beq{shsis}
\gmp(\th)= {\cos(2\alpha) \over |\th-\u|} \ \ \ \ \
\gmc(\th)= {\sin(2\alpha)\over |\th-\u|}. 
\eeq
The cosine and sine of  $2 \alpha$ can be calculated in terms of
\beqa{al}
\label{cosal}
\cos(\alpha)&=&{(\th-\u) \cdot (\th-\o) \over |\th-\u| \ |\th-\o|}  \\
\label{sineal}
\sin(\alpha)&=&{\hat{z} \cdot [(\th-\u) \times(\th-\o)] \over |\th-\u| \ |\th-\o|} ,
\eeqa
where $\hat{z}$ is the unit vector perpendicular to the plane of the
sky. The three-point function $\zeta$ is obtained by integrating over
the position of the center of the halo $\u$,
\beqa{sisint}
\label{shearcos}
\zeta_\gamma^{+++}(l_1,l_2,l_3)&=&\int d^2\u {\cos(2\alpha_1) \over |\th_1-\u|} {\cos(2\alpha_2) \over |\th_2-\u|} {\cos(2\alpha_3) \over |\th_3-\u|}   \\ \nonumber \\
\label{shearsin}
\zeta_\gamma^{+\times \times}(l_1,l_2,l_3)&=&\int d^2\u {\cos(2\alpha_1) \over |\th_1-\u|} {\sin(2\alpha_2) \over |\th_2-\u|} {\sin(2\alpha_3) \over |\th_3-\u|},
\eeqa
where $l_1,l_2,l_3$ are the lengths of the triangle sides,
$l_1^2=|\th_2-\th_1|^2$, $l_2^2=|\th_3-\th_2|^2$,
$l_3^2=|\th_1-\th_3|^2$. There are two permutations of the second
equation which give the remaining three-point correlators,
$\zeta_\gamma^{\times + \times}$ and $\zeta_\gamma^{\times \times +}$.

\begin{figure}[t]
\centerline{\epsfxsize=9cm\epsffile{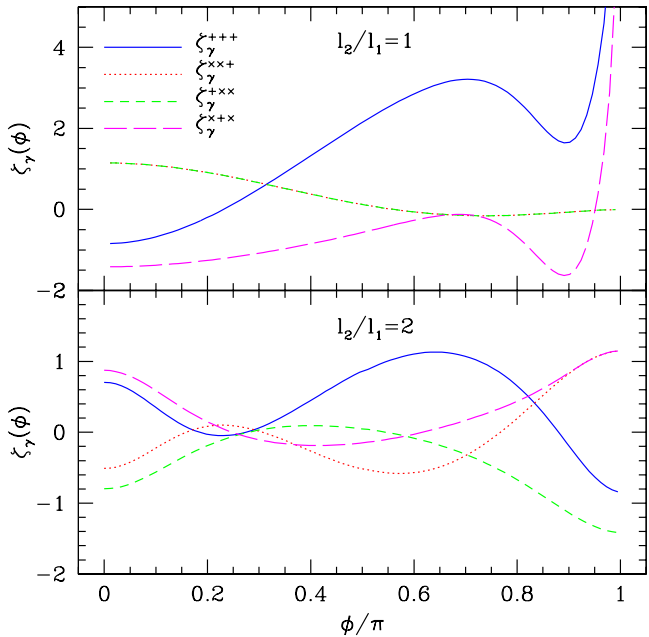}}
\caption{The shear three-point functions $\zeta_\gamma(l_1,l_2,l_3)$ for the case of isothermal sphere halos, as a function of the angle $\phi$ between ${\bf l}_1$ and ${\bf l}_2$ for $l_2=l_1$ (top panel) and $l_2=2l_1$ (bottom panel). Different line styles correspond to the four shear correlators, as labeled in the top panel.}
\label{shearSIS}
\end{figure}

To calculate Eqs.~(\ref{shearcos}-\ref{shearsin}), we proceed as
follows. First, for simplicity, we take the origin of coordinates to
coincide with the center of mass, so $\o$ vanishes. We then write the
$\u$ dependence inside the cosine and sine in
Eqs.~(\ref{cosal}-\ref{sineal}) in terms of ${\bf d}_i \equiv
\th_i-\u$. This can be done simply using that

\beq{qaz1}
\u \cdot \th_i = \frac{1}{2} \Big(u^2+|\th_i|^2 -d_i^2\Big),
\eeq
whereas the magnitude of $\u$ can be conveniently written as
\beq{qaz2}
u^2= {1\over 3} (d_1^2+d_2^2+d_3^2) -{1\over 9} (l_1^2+l_2^2+l_3^2) .
\eeq
Similarly we have  $|\th_1|^2 =(2l_3^2+2l_1^2-l_2^2)/9$ and cyclic
permutations. In this way, after appropriate translation in $\u$ the
integrals in Eqs.~(\ref{shearcos}-\ref{shearsin}) are of the form,

\beq{jdef}
J(\nu_1,\nu_2,\nu_3) =
\int  \frac{d^2 {\bf u}}{(u^2)^{\nu_1} [({\bf l_{1}}-{\bf u})^2]^{\nu_2}
[({\bf l_{2}}-{\bf u})^2]^{\nu_3}} 
\eeq
which can be evaluated by using dimensional regularization techniques
(see Scoccimarro 1997 and references therein) in terms of Apell's
hypergeometric function of two variables, ${\rm F}_{4}$, with the
series expansion:
\begin{equation}
	{\rm F}_{4} (a,b;c,d;x,y) = \sum_{i=0}^{\infty} \sum_{j=0}^{\infty}
     \frac{x^{i}y^{j}}{i!\ j!} \ \frac{ (a)_{i+j} (b)_{i+j}}{(c)_{i}
     (d)_{j}}
	\label{F4exp},
\end{equation}
where $(a)_{i} \equiv \Gamma (a+i) / \Gamma (a)$. In our case, the
arguments of $F_4$ have a special symmetry that allows us to write
them in terms of regular hypergeometric functions,
\beqa{f4hy}
F_4[\alpha,\gamma+\gamma^\prime-\alpha-1,\gamma,\gamma^\prime;x(x-1),y(y-1)] = 
\nonumber \\ F(\alpha,\gamma+\gamma^\prime-\alpha-1,\gamma,x)\times  F(\alpha,\gamma+\gamma^\prime-\alpha-1,\gamma^\prime,y), \nonumber \\
\eeqa
which can be easily evaluated on the computer using {\small
MATHEMATICA}. Figure~\ref{shearSIS} shows the results for different
triangle configurations. Since the SIS is scale-free, the overall size
of the triangle scales the results by a factor, so we take
$l_1\equiv 1$. The top panel shows the four $\zeta_\gamma$'s for
$l_2=l_1$ as a function of the angle $\phi$ between ${\bf l}_1$ and
${\bf l}_2$. We see that for most of the triangles ($\phi \gtrsim
0.3\pi$), the signal is dominated by $\zeta_\gamma^{+++}$, with a
maximum close to equilateral triangles ($\phi=2\pi/3$)\footnote{The
divergence of $\zeta_\gamma^{+++}$ as $\phi \rightarrow \pi$ is a
peculiarity of the SIS when $l_3 \rightarrow
0$. $\zeta_\gamma^{\times+\times}$ is regular in that limit despite
what Fig.~\protect\ref{shearSIS} suggests.}. On the other hand, the
three contributions involving $\gmc$ are generally smaller in
magnitude.

\begin{figure}[t]
\centerline{\epsfxsize=9cm\epsffile{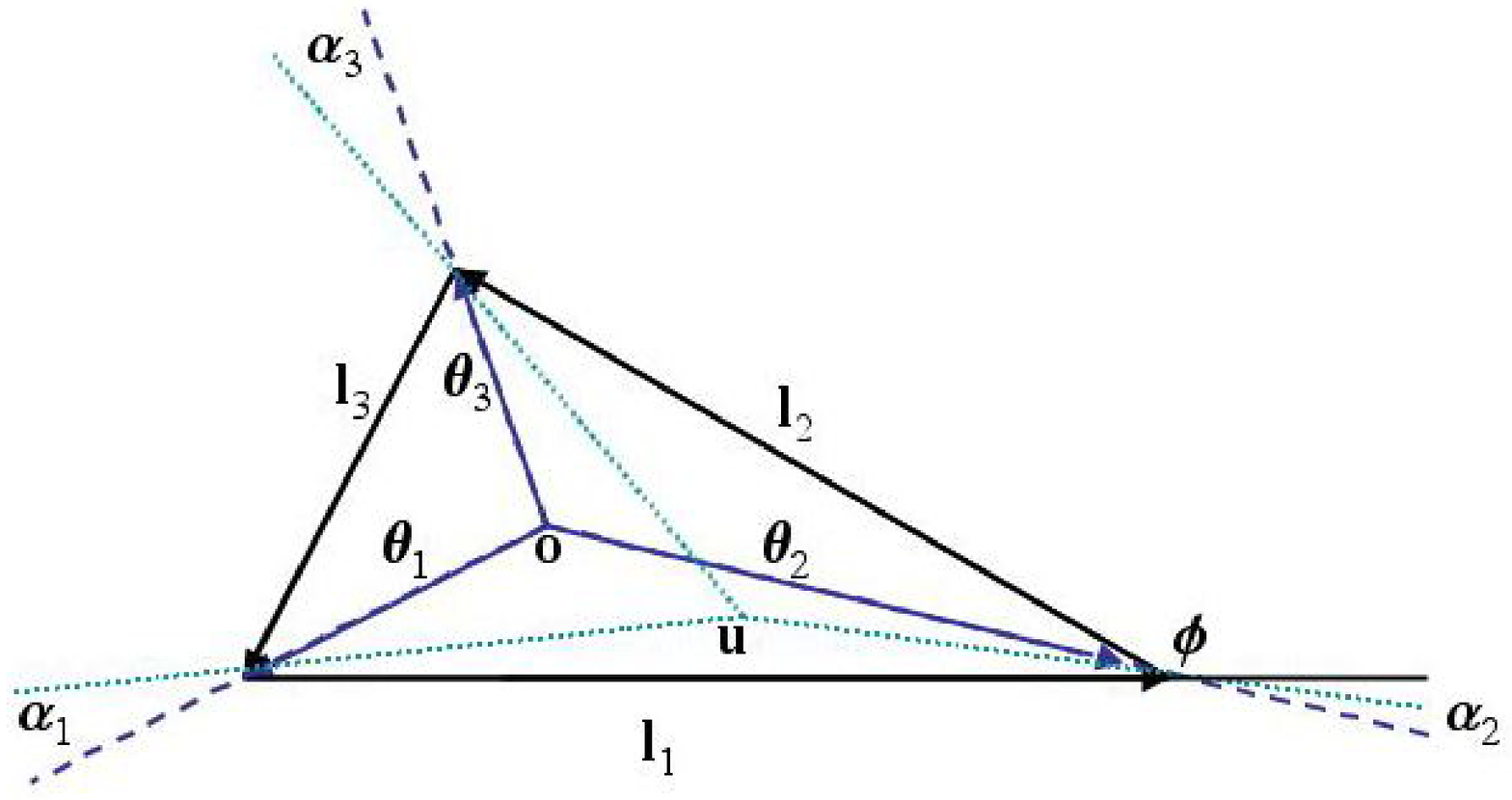}}
\caption{Definition of triangle variables. We characterize the shear
generated by a halo with center at $\u$ by measuring its components
with respect to the center of mass of the
triangle, denoted by $\o$. }
\label{tmaestro}
\end{figure}

This can be understood geometrically from Fig.~\ref{tmaestro}. Let us
consider the halo center to be inside the triangle, which minimizes
the distances to the vertices and thus maximizes the signal. Since
$\cos (2\alpha)>0$ for $|\alpha|<\pi/4$, as long as the internal
angles of the triangle are smaller than $\pi/2$, $\zeta_\gamma^{+++}$
is positive. As $\phi \rightarrow \pi$, $\alpha_2 \rightarrow 0$,
whereas $|\alpha_1|=|\alpha_3|$ cannot be larger than $\pi/2$; thus
$\zeta_\gamma^{+++}$ remains positive as $\phi \rightarrow \pi$. On
the other hand, as $\phi \rightarrow 0$, $\alpha_1=\alpha_3
\rightarrow 0$, whereas $|\alpha_2| \rightarrow \pi$ if $\u$ is off
center; that explains why $\zeta_\gamma^{+++}$ becomes negative as
$\phi \rightarrow 0$.

For the three other three-point functions involving $\gmc$, things are
more subtle. Since $\sin (2\alpha)$ changes sign at $\alpha=0$, $\gmc$
can be positive or negative depending on the location of $\u$ relative
to the bisector of the vertex. As a result moving $\u$ inside the
triangle leads to cancellations, and thus smaller amplitude for
$\zeta_\gamma$'s. The bottom panel in Fig.~\ref{shearSIS} shows
results for configurations in which $l_2=2l_1$. A similar pattern is
seen, where for isosceles triangles $\phi \approx 0.6 \pi$ the
$\zeta_\gamma^{+++}$ is positive and maximum, whereas the remaining
three contributions are smaller, becoming comparable only for
collapsed triangles which can avoid cancellations.

\begin{figure}[h]
\centerline{\epsfxsize=9cm\epsffile{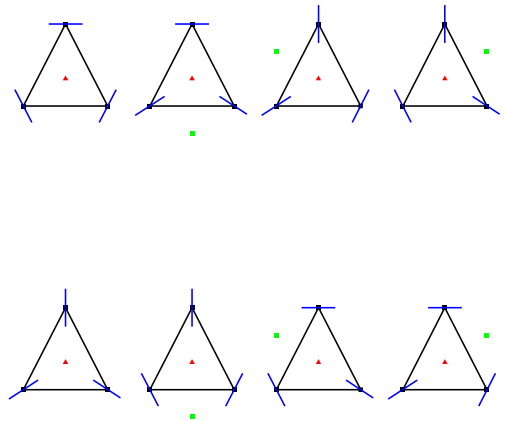}}
\caption{Examples of shear patterns generated at the vertices of
equilateral triangles by overdensities (top panel) and underdensities
(bottom panel) located at the position denoted by square symbols. From
left to right, contributions to $\zeta_\gamma^{+++}$,
$\zeta_\gamma^{\times\times+}$, $\zeta_\gamma^{\times +\times}$, and
$\zeta_\gamma^{+\times\times}$ (vertices labeled anticlockwise
starting from bottom-left vertex). Note that on the left-most plots,
the density perturbation is located at the center of the triangle.}
\label{cosine}
\end{figure}

The physical interpretation of the different amplitudes is sketched in
Fig.~\ref{cosine}. In the top (bottom) row we show patterns that would produce 
positive (negative) values for $\zeta_\gamma^{+++}$, $\zeta_\gamma^{\times\times+}$,
$\zeta_\gamma^{\times +\times}$, and $\zeta_\gamma^{+\times\times}$
from left to right, respectively. It is clear from the figure that the
top examples correspond to patters that would be produced by
overdensities and the bottom patterns would be produced by
underdensities. We indicate with a square the region where the
overdensity (underdensity) should be located to produce such
pattern. 

\subsection{Superposition of NFW profiles}

In this section we will calculate the three-point functions using the
halo model (Peacock \& Smith 2000, Seljak 2000, Ma \& Fry 2000,
Scoccimarro et al. 2001). For simplicity we will restrict ourselves to
the one-halo term which dominates on the small scales where the
three-point function of the shear is easiest to measure in
observations. For examples of calculations of weak gravitational
lensing higher-order moments in the context of the halo model see
\cite{hucoo} and \cite{takdajain}. Measurements of higher-order
moments of the convergence field in numerical simulations are given in
Jain, Seljak \& White (2000), White \& Hu (2000); Van Waerbeke et al
(2001) also present results for aperture mass statistics which is
directly related to the cosmic shear.

Under our assumptions the averaged shear can be written as an integral
over radial distance ($d \chi = c\ dt/a$), mass ($M$) and angular
location ($\u$) of halos,
\beqa{halo3pt}
\zeta_\gamma(l_1,l_2,l_3)&=&\int d\chi  \ d_A^2(\chi) \int dM    {dn\over dM}  
\int d^2\u \nonumber \\    \tlg(\u,\th_1,M,\chi) && \tlg(\u,\th_2,M,\chi) \ \ \tlg(\u,\th_3,M,\chi),
\eeqa
where $a$ is the expansion factor of the universe, $d_A$ is the
comoving angular diameter distance, $ {dn/dM}$ is the mass function of
halos.  We have introduced the notation $\tlg(\u,\th,M,\chi) $ to
indicate the shear produced at position $\th$ by a halo at radial
distance $\chi$ and angular position $\u$. Note that for convenience 
we are using the same symbol regardless 
of whether $\gmp$ or $\gmc$ is involved.

To obtain an estimate for the three-point function we will evaluate
Eq.~(\ref{halo3pt}) assuming that the background sources used to
measure the shear are all at redshift $z_s=1$ and that the
cosmological model is the so called LCDM model ( $\Omega_m=0.3$,
$\Omega_\Lambda=0.7$, $h=0.7$, $\sigma_8=0.9$ and $n=1$). We will assume
that the mass function of halos is that given by \cite{shto,jenk} and
that dark matter halos have an NFW profile \cite{nfw}. In particular,
we use
\beq{nfwp}
\rho(r)={\rho_s \over r/r_s (1+r/r_s)^2}
\eeq
where $r$ is measured in comoving coordinates and $r_s$ is related to
the virial radius of the halo by the concentration parameter $c$,
$r_s=r_{vir}/c$. The mass of the halo is given by $M=4\pi \rho_s
r_{vir}^3(\ln(1+c)-c/(1+c))/c^3$. The virial radius is calculated
using that $M=4\pi/3 r_{vir}^3 \bar \rho_0 \Delta(z)$ with $\bar
\rho_0$ the mean density of the universe today and $\Delta(z)$ the
overdensity of collapse as a function of redshift (ie.  $\Delta(z=0)
\approx 340$ for LCDM). For the concentration we take $c(M,z)=9/(1+z)
(M/M_*)^{-0.13}$ \cite{bullock}, where $M_*$ is the mass contained in
a sphere of radius $R_*$ at which the variance of the density is one
($R_* \approx 3.14\ h^{-1} {\rm Mpc}$ for LCDM).

The shear produced by an NFW profile at the origin is calculated as
follows. The shear is expressed in terms of second derivatives of the
gravitational potential $\psi$, $\gamma_1=1/2 (\psi_{xx} -\psi_{yy})$
and $\gamma_2=\psi_{xy}$. The gravitational potential satisfies
\beqa{poten}
\nabla^2_{\u}\psi&=& 2 \kappa(u) \nonumber \\
\kappa(u) &=&  {1\over a \Sigma_{\rm crit} } \int dz \ \rho(\sqrt{u^2 d^2_A(\chi)+z^2})
\eeqa   
where $1/\Sigma{\rm crit} = 4\pi G d_A(\chi) d_A(\chi-\chi_s)/c^2
d_A(\chi_s)$, $\chi$ gives the radial position of the halo and
$\chi_s$ that of the background sources\footnote{In this equation $c$
is the speed of light not the concentration parameter of dark matter
halos}.  Equation (\ref{poten}) is easily integrated to obtain $\psi$
because $\kappa$ is only a function of $u$.  Once the shear is
obtained for a halo at the origin, we obtain $\tlg(\u,\th,M,\chi)$
with a coordinate transformation. A similar evaluation for the SIS to
compare with the results obtained by the method of the previous
section gives a useful check to our numerical integration code.

\begin{figure}[tb]
\centerline{\epsfxsize=9cm\epsffile{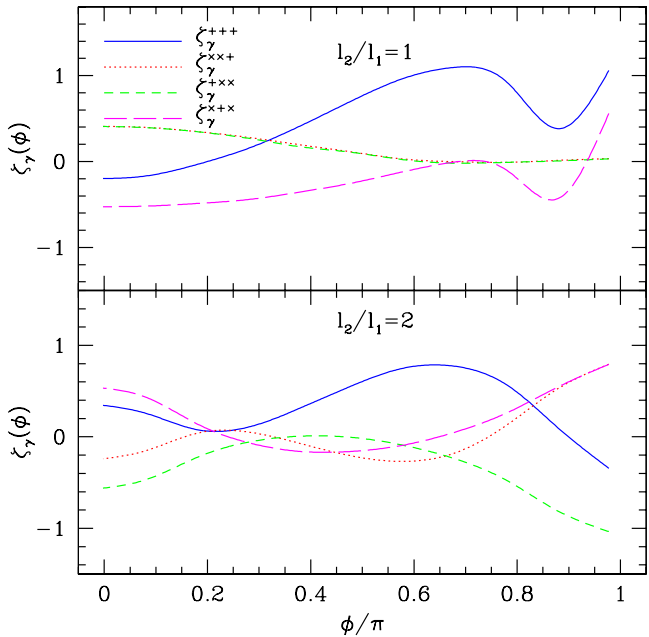}}
\caption{Even components of the three-point function of the shear
calculated using the 1-halo term for LCDM. The top (bottom) panel
shown the case $l_1=l_2=2'$ ($l_1=2'$, $l_2=4'$). }
\label{fignfw}
\end{figure}

Figure \ref{fignfw} shows the results of our calculation for some
specific triangles. In the top panel $l_1=l_2=2 '$ and in the bottom
panel the case $l_1=2'$ and $l_2=4'$. The similarities with the
results for the SIS are striking, thus we can understand the
dependence of each of the curves with $\phi$ in exactly the same way.

The expected level of $\zeta_\gamma$ on arcminute scales
is $\zeta_\gamma \sim 10^{-6}$. We  estimate the number of
triangles ($N_T$) needed to detect the three-point function in a
particular configuration above the noise produced by the intrinsic
ellipticity of the background galaxies as follows. If the typical
ellipticity of the background galaxies is $\sim 0.3$ then the typical
noise added to each components of the intrinsic shear is $\sigma\sim
0.3/\sqrt2 \sim 0.2$. The
expected noise in the three-point function is then $\sigma^3/\sqrt
N_T$. This implies that one needs roughly $N_T \sim 6.5\times 10^7$ triangles of
a particular shape to estimate the three-point function. In a survey
covering a solid angle $\Omega$ with a mean desity of galaxies $\bar n
$ there are roughly $N_T=\bar n \Omega \ (\pi R^2 \bar n)^2/6$
triangles with sides of scale $R$. Current surveys have about 
20 galaxies/${\rm arcmin}^2$ implying that in a few square degrees there 
are enough triangles with sides of order an arcminute 
to detect this signal. 

\section{Summary and Discussion}

In this paper we have introduced a new way of defining higher-order
correlation functions of a spin-2 field such as the weak lensing shear
or the CMB polarization by using the ``center of mass" of the
configuration as the origin from which the components of the shear are
defined. In principle for an $N$-point function there are $2^N$
different statistics of the shear.  We have shown that these
statistics can be divided according to their behavior under parity
transformation and that one does not expect a cosmological signal in
the odd ones for some configurations, such as for equilateral triangles.

In order to gain intuition about the behavior of these statistics we
calculated the four even three-point functions under some simple
assumptions. We calculated analytically what would be expected to be
produced by an ensemble of singular isothermal spheres. We showed that
$\zeta_\gamma^{+++}$ is positive and expected to carry the bulk of the
signal for triangles that are not too elongated. For elongated
triangles we showed that all four three-point functions have similar values
but some of them are positive and others are negative.  If two of the
points are very close to each another $\zeta_\gamma^{+++}$ and
$\zeta_\gamma^{\times+\times}$ carry most of the signal.

We estimated the three-point functions in the context of the halo
model using the contributions from the one-halo term, which
should be a reasonable approximation at small angular scales. We
showed that the configuration dependence in this case is almost
identical to that found for the SIS. We estimated that in order to 
detect a signal above the noise (due to the intrinsic ellipticity
of galaxies) at scales of order of one arcminute a survey 
of a few square degrees is necessary. 

Clearly both the fact that we restricted ourselves to the one-halo
term and that we assumed the halos to be spherical will affect the
behavior of the three-point function. A detailed study using numerical
simulations will be needed to improve upon the calculation presented
here (Benabed, Scoccimarro \& Zaldarriaga, in preparation).  The
potential rewards of detecting a non-Gaussian signal in the shear maps
are enourmous and the tantalizing detections reported so far
(\cite{BMVWb}) make this a very exciting time to study these
issues in detail.

\vskip 0.5cm 
When this paper was under completion, a similar proposal
for calculating the shear three-point function was put forward by
Schneider \& Lombardi (2002).
\vskip 0.5cm

Acknowledgments: We thank Mashiro Takada and Bhuvnesh Jain for
pointing out an error in the first version of the manuscript. 
MZ is supported by David and Lucille Packard
Foundation Fellowship for Science and Engeneering and NSF grant
AST-0098506. RS is supported by NASA ATP grant NAG5-12100. MZ and RS
are supported by NSF grant PHY-0116590.

\end{document}